\documentclass[12pt]{article}
\usepackage{cite}
\usepackage{amsmath}
\begin{document}

\title{\hfill{\normalsize{}hep-th/0603212}\\[5mm]
{\bf{}Gauge invariant Lagrangian construction for
massive higher spin fermionic fields}}

\author{\sc I.L. Buchbinder${}^{a}$\thanks{\tt joseph@tspu.edu.ru},
V.A. Krykhtin${}^b$\thanks{krykhtin@mph.phtd.tpu.edu.ru},
L.L. Ryskina$^a$\thanks{riskinall@tspu.edu.ru},
H. Takata$^a$\thanks{takataxx@tspu.edu.ru}
\\[0.5cm]
\it ${}^a$Department of Theoretical Physics,\\
\it Tomsk State Pedagogical University,\\
\it Tomsk 634041, Russia\\[0.3cm]
\it ${}^b$Laboratory of Mathematical Physics and\\
\it Department of Theoretical and Experimental Physics\\
\it Tomsk Polytechnic University,\\
\it Tomsk 634050, Russia}

\date{}

\maketitle
\thispagestyle{empty}

\begin{abstract}
We formulate a general gauge invariant Lagrangian construction
describing the dynamics of massive higher spin fermionic fields in
arbitrary dimensions. Treating the conditions determining the
irreducible representations of Poincare group with given spin as
the operator constraints in auxiliary Fock space, we built the
BRST charge for the model under consideration and find the gauge
invariant equations of motion in terms of vectors and operators in
the Fock space. It is shown that like in massless case
\cite{0410215}, the massive fermionic higher spin field models are
the reducible gauge theories and the order of reducibility grows
with the value of spin. In compare with all previous approaches,
no off-shell constraints on the fields and the gauge parameters
are imposed from the very beginning, all correct constraints
emerge automatically as the consequences of the equations of
motion. As an example, we derive a gauge invariant Lagrangian for
massive spin $3/2$ field.

\end{abstract}


\section{Introduction}

Higher spin field theory is one of the actively developing trends of
modern theoretical physics. The various approaches to higher spin
problem and current results are widely discussed in the literature
(see e.g.  \cite{reviews} for reviews and \cite{devm,dev0} for recent
developments for massive and massless case respectively). In this note
we extend the new gauge invariant approach, developed in the papers
\cite{0410215,0505092,0511276} for massless higher spin fermionic
fields and for massive higher spin bosonic fields (see discussion of
motivations and features of this approach in \cite{0505092}), to the
massive higher spin fermionic field models.

The main idea of the approach under consideration is treatment of
the conditions determining the irreducible representation of the
Poincare group with given spin in terms of operator constraints in
some auxiliary Fock space and use of the BRST-BFV construction
\cite{BRST} to restore a Lagrangian of the theory on the base of
given system of constraints. 
Such an approach was firstly realized in open string theory \cite{Witten} 
and then applied to higher spin field theory \cite{BRST-HS}. Our aim
is to extend the recent results obtained in \cite{0410215,0505092} 
for massless fermionic and and massive bosonic higher spin field theories
respectively to massive fermionic theories.

The massive fermionic higher spin theories have the specific
differences with the massive bosonic ones \cite{0505092} and
demand a separate consideration. In contrast to the bosonic case
and similar to the massless fermionic case \cite{0410215} we get
the fermionic operators in the algebra of constraints and
corresponding them the bosonic ghosts.  Due to the presence of the
bosonic ghosts the resulting theory is a gauge theory where an
order of reducibility grows with the spin of the field.  Except
the presence of bosonic ghosts in the fermionic theory there is
one more special aspect. Unlike the bosonic case, in the fermionic
theory we have two Hermitian constraints thus there is a problem
in constructing an appropriate scalar product. It means, we can
not obtain a consistent Lagrangian for fermions by the same method
which was used in the bosonic theory since such a method
automatically leads to second order equations of motion for
fermionic fields what contradicts to the spin-statistic theorem.
To overcome this problem we partially fix the gauge and partially
solve some field equations removing the ghost fields corresponding
to the Hermitian constraints and thus derive the correct
Lagrangian.

The paper is organized as follows. In section \ref{ac} we
introduce the  operator algebra generated by primary constraints
which define irreducible representation of the Poincare group with
fixed arbitrary half-integer spin. In section~\ref{newrep} we
formulate a new representation of the operators and a
corresponding new scalar product which overcome a problem arising
at naive use of BRST construction (see a discussion of this
problem in \cite{0410215}, \cite{0505092}). Then, in
section~\ref{Lagr} we construct the BRST operator and derive the
Lagrangian for a massive field with given half-integer spin, the
corresponding gauge transformations and equations of motion in
terms of vectors in Fock space. In section~\ref{example} we
illustrate the procedure of Lagrangian construction by finding the
gauge Lagrangian for massive field with spin $3/2$.
Section~\ref{Summary} summarizes the results obtained.

\section{Algebra of the constraints}\label{ac}

It is well known that the totally symmetrical tensor-spinor fields
$\Phi_{\mu_1\cdots\mu_n}$ (the Dirac index is suppressed),
describing the irreducible massive spin $s=n+1/2$ representations
of the Poincare group must satisfy the following conditions (see
e.g. \cite{BK})
\begin{eqnarray} && (i\gamma^\nu\partial_\nu-m)
\Phi_{\mu_1\cdots\mu_n}=0,
\label{irrep0}
\\
&&
\gamma^\mu\Phi_{\mu\mu_2\cdots\mu_n}=0.
\label{irrep1}
\end{eqnarray}
Here $\gamma^\mu$ are the Dirac matrices
\begin{eqnarray}
\{\gamma_\mu,\gamma_\nu\}=2\eta_{\mu\nu},
&\qquad&
\eta_{\mu\nu}=(+,-,\ldots,-).
\label{Diracgamma}
\end{eqnarray}

To avoid a manipulations with a number of indices it is convenient to
introduce auxiliary Fock space generated by creation and annihilation
operators $a_\mu^+$, $a_\mu$ with vector Lorentz index
$\mu=0,1,2,\ldots,D-1$ satisfying the commutation relations
\begin{eqnarray}
\bigl[a_\mu,a_\nu^+\bigr]=-\eta_{\mu\nu}.
\end{eqnarray}
These operators act on vectors in the
Fock space
\begin{eqnarray}
|\Phi\rangle&=&\sum_{n=0}^{\infty}\Phi_{\mu_1\cdots\mu_n}(x)
a^{+\mu_1}\cdots a^{+\mu_n}|0\rangle
\label{gstate}
\end{eqnarray}
These vectors are the functionals of the higher spin fields and
can be called the higher spin functionals. The conditions
(\ref{irrep0}), (\ref{irrep1}) are realized as the constraints on
the vectors (\ref{gstate}) in the form
\begin{eqnarray}
T_0^{\,\prime}|\Phi\rangle=0, &\qquad& T_1^{\,\prime}|\Phi\rangle=0,
\label{01'}
\end{eqnarray}
where
\begin{eqnarray}
\label{T0'}
T_0^{\,\prime}&=&\gamma^\mu{}p_\mu+m,
\\
T_1^{\,\prime}&=&\gamma^\mu{}a_\mu
,
\label{T1'}
\end{eqnarray}
with $p_\mu=-i\frac{\partial}{\partial{}x^\mu}$.
If
constraints (\ref{01'}) are fulfilled for the general state
(\ref{gstate}) then conditions (\ref{irrep0}), (\ref{irrep1})
are fulfilled for each component $\Phi_{\mu_1\cdots\mu_n}(x)$ in
(\ref{gstate}) and hence vectors (\ref{gstate}) under relations
(\ref{01'}) describe the free arbitrary higher spin fermionic fields.

Important element of general BRST construction is finding a closed
algebra of all constraints on the field $|\Phi\rangle$
(\ref{gstate}) generated by their (anti)commutators.  Finding such
an algebra in the case under consideration is similar to that in
the massless fermionic case \cite{0410215}.  In the massless case
the constraints which are analog of $T_0^{\,\prime}$,
$T_1^{\,\prime}$ should be treated as fermionic.  In the massive
case the constraint $T_0^{\,\prime}$ (\ref{T0'}) looks like a sum
of two terms: one is fermionic massless Dirac operator
$i\gamma^\mu{}p_\mu$ like in the massless case and another one is
bosonic mass operator $m$.  This situation looks contradictory. In
order to avoid the contradiction ones act as follows.  We
introduce another constraints $T_0$, $T_1$ on the state vector
(\ref{gstate}) instead of $T_0^{\,\prime}$, $T_1^{\,\prime}$
\begin{eqnarray}
T_0&=&\tilde{\gamma}{}^\mu p_\mu+\tilde{\gamma}m , \label{T0} \\
T_1&=&\tilde{\gamma}{}^\mu a_\mu , \label{T1}
\end{eqnarray}
where
$\tilde{\gamma}{}^\mu$, $\tilde{\gamma}$ is a set of $D+1$ Grassmann
odd objects which obey the following gamma-matrix-like conditions
\begin{eqnarray}
\{\tilde{\gamma}{}^\mu,\tilde{\gamma}{}^\nu\}=2\eta^{\mu\nu},
&\qquad&
\{\tilde{\gamma}{}^\mu,\tilde{\gamma}\}=0,
\qquad
\tilde{\gamma}{}^2=-1.
\label{tildedgamma}
\end{eqnarray}
We have no necessity to realize the above objects in explicit form
since for construction of the final Lagrangians we will switch
from "gamma-matrix-like objects" (\ref{tildedgamma}) to the "true"
gamma-matrices by the relation
\begin{eqnarray}
\gamma^\mu&=&\tilde{\gamma}{}^\mu\tilde{\gamma}
           =-\tilde{\gamma}\tilde{\gamma}{}^\mu.
\label{truegamma}
\end{eqnarray}
One can check that the relations (\ref{tildedgamma}) lead to the
relations (\ref{Diracgamma}) for "true" gamma-matrices
(\ref{truegamma}).
Then ones can show that $T_0$, $T_1$ and $T_0^{\,\prime}$,
$T_1^{\,\prime}$ related by the following conditions
\begin{eqnarray}
T_0=\tilde{\gamma}T_0^{\,\prime},
&\qquad&
T_1=\tilde{\gamma}T_1^{\,\prime}.
\end{eqnarray}
Thus equations (\ref{01'}) are equivalent to the equations
\begin{eqnarray}
T_0|\Phi\rangle=0,
&\qquad&
T_1|\Phi\rangle=0
\label{01}
\end{eqnarray}
and therefore we can treat just the operators $T_0$ (\ref{T0}) and $T_1$
(\ref{T1}) as primary constraints.
Commutators of these constraints generate all other constraints on the
ket-vector state (\ref{gstate}).  Thus we get three more
constraints\footnote{Algebra of the operators will be written down in
an explicit form later. See Table~\ref{table} at page~\pageref{table}.}
\begin{eqnarray}
L_0|\Phi\rangle=
L_1|\Phi\rangle=
L_2|\Phi\rangle=0,
\end{eqnarray}
where
\begin{eqnarray}
L_0=-p^2+m^2,
\qquad
L_1=a^\mu p_\mu,
\qquad
L_2={\textstyle\frac{1}{2}}\,a_\mu a^\mu
.
\label{L}
\end{eqnarray}

Our purpose is to construct the Lagrangian for the massive fermionic
higher spin fields generalizing the
method of \cite{0410215,0505092,0511276}.
This method is based on finding an Hermitian BRST operator
on the base of some system of operator constraints forming a closed
algebra invariant under Hermitian conjugation.
Hence, in our case we should look for a set of operator constraints
which is invariant under Hermitian conjugation.  Let us define a scalar
product in the Fock space taking bra-vector state in the form
\begin{eqnarray}
\langle\tilde{\Phi}|&=& \sum_{n=0}^{\infty}\langle0|a^{\mu_1}\ldots
a^{\mu_n} \Phi^+_{\mu_1\ldots\mu_n}(x)\tilde{\gamma}{}^0
\label{tildedstate}
\end{eqnarray}
and postulate the following Hermitian
properties\footnote{It can be checked that relations
(\ref{tildedH}) lead to usual hermitian relations for the "true"
gamma-matrices (\ref{truegamma})
\begin{eqnarray}
(\gamma^\mu)^+
=
\gamma^0\gamma^\mu\gamma^0.
\end{eqnarray}
}
for
$\tilde{\gamma}{}^\mu$ and $\tilde{\gamma}$
\begin{eqnarray}
\label{tildedH}
(\tilde{\gamma}{}^\mu)^+
=
\tilde{\gamma}{}^0\tilde{\gamma}{}^\mu\tilde{\gamma}{}^0,
&\qquad&
(\tilde{\gamma})^+
=
\tilde{\gamma}{}^0\tilde{\gamma}\tilde{\gamma}{}^0
=
-\tilde{\gamma}
.
\end{eqnarray}
As a result we get two Hermitian constraints $T_0$ and $L_0$ and
the three other are non Hermitian\footnote{In what follows we
shall write $F^+=(F)^+$ instead of
$F^+=\tilde{\gamma}{}^0(F)^+\tilde{\gamma}{}^0$ and omit
$\tilde{\gamma}{}^0$ in writing hermitian conjugation}
\begin{eqnarray}
T_1^+=
\tilde{\gamma}{}^0(T_1)^+\tilde{\gamma}{}^0=
\tilde{\gamma}{}^\mu a_\mu^+,
\qquad
L_1^+=(L_1)^+=a^{+\mu}p_\mu,
\qquad
L_2^+=(L_2)^+={\textstyle\frac{1}{2}}\,a^{+\mu}a_\mu^+.
\label{L+}
\end{eqnarray}
Therefore we extend the set of constraints adding three new
operators $T_1^+$, $L_1^+$, $L_2^+$ to the initial constraints
(\ref{T0}), (\ref{T1}), (\ref{L}) on the ket-state vector
(\ref{gstate}).
As a result a set of the constraints
$T_0$, $T_1$, $T_1^+$, $L_0$, $L_1$, $L_1^+$, $L_2$, $L_2^+$
is invariant under Hermitian conjugation.

Algebra of the operators $T_0$, $T_1$, $T_1^+$, $L_0$, $L_1$,
$L_1^+$, $L_2$, $L_2^+$ is open in terms of (anti)commutators of
these operators. According to \cite{0410215,0505092,0511276} we
must include in the set of operators all the operators which are
needed for the algebra be closed. These operators are
\begin{eqnarray}
G_0=-a_\mu^+a^\mu+{\textstyle\frac{D}{2}},
\qquad
M_1=\tilde{\gamma}\,m,
\qquad
M_2=-m^2.
\label{nonconstr}
\end{eqnarray}
Since these operators are obtained as (anti)commutators of a constraint
on the ket-vector state (\ref{gstate}) with a constraint on
bra-vector state (\ref{tildedstate}) then operators (\ref{nonconstr})
can't be considered as constraints neither on the space of
bra-vectors nor on the space of ket-vectors.
Total algebra of the operators is written in Table~\ref{table},
\begin{table}[t]
\small
\begin{eqnarray*}
\begin{array}{||l||r|r|r|r|r|r|r|r||r|r|r||}\hline\hline\vphantom{\Biggm|}
&T_0&T_1&T_1^+&\quad{}L_0&L_1&L_1^+&L_2&L_2^+&G_0&M_1&M_2\\
\hline\hline\vphantom{\Biggm|}
T_0
   &-2L_0&2L_1&2L_1^+&0&0&0&0&0&0&2M_2&0\\
\hline\vphantom{\Biggm|}
T_1
   &2L_1&4L_2&-2G_0&0&0&-(\ref{L1+T1})&0&-T_1^+&T_1&0&0\\
\hline\vphantom{\Biggm|}
T_1^+
   &2L_1^+&-2G_0&4L_2^+&0&(\ref{L1+T1})&0&T_1&0&-T_1^+&0&0\\
\hline\vphantom{\Biggm|}
L_0
   &0&0&0&0&0&0&0&0&0&0&0\\
\hline\vphantom{\Biggm|}
L_1
   &0&0&-(\ref{L1+T1})&0&0&(\ref{L1L1+})&0&-L_1^+&L_1&0&0\\
\hline\vphantom{\Biggm|}
L_1^+
   &0&(\ref{L1+T1})&0&0&-(\ref{L1L1+})&0&L_1&0&-L_1^+&0&0\\
\hline\vphantom{\Biggm|}
L_2
   &0&0&-T_1&0&0&-L_1&0&G_0&2L_2&0&0\\
\hline\vphantom{\Biggm|}
L_2^+
   &0&T_1^+&0&0&L_1^+&0&-G_0&0&-2L_2^+&0&0\\
\hline\hline\vphantom{\Biggm|}
G_0
   &0&-T_1&T_1^+&0&-L_1&L_1^+&-2L_2&2L_2^+&0&0&0\\
\hline\vphantom{\Biggm|}
M_1
   &2M_2&0&0&0&0&0&0&0&0&2M_2&0\\
\hline\vphantom{\Biggm|}
M_2
   &0&0&0&0&0&0&0&0&0&0&0\\
\hline\hline
\end{array}
\end{eqnarray*}
\caption{Algebra of the operators}\label{table}
\end{table}
where
\begin{eqnarray}
[L_1^+,T_1]&=&[T_1^+,L_1]=T_0-M_1
\label{L1+T1}
,
\\{}
[L_1,L_1^+]&=&L_0+M_2
\label{L1L1+}
.
\end{eqnarray}

The operators $T_0$, $T_1$, $T_1^+$, $M_1$
are fermionic and the
operators $L_0$, $L_1$, $L_1^+$, $L_2$, $L_2^+$, $G_0$, $M_2$ are
bosonic.
All the commutators are graded, i.e. graded commutators between
the fermionic operators are anticommutators and graded
commutators which include any bosonic operator are
commutators.
In Table~\ref{table} the first arguments of the graded
commutators are
listed in the left column and the second argument of graded
commutators are listed in the upper row. We will call this algebra as
free massive half-integer higher spin symmetry algebra.

Method of deriving the higher spin Lagrangians on the base of
operator algebra  analogous one given by Table
~\ref{table} was discussed in \cite{0410215} and
\cite{0505092}. One can show that a naive use of BRST
construction leads to some contradictions (see
discussions of this point in \cite{0410215,0505092}) and is
directly applicable only for spin 1/2 case.  According to the method
developed in \cite{0410215,0505092,0511276} to overcome
these contradictions we should pass to new representation for the
operator algebra given in Table~\ref{table}.  This will be done in the
next section.

\section{New representation of the algebra}\label{newrep}

The approach developed in
\cite{0410215,0505092,0511276} is based on special representation of
the operator algebra generated by constraints. In the case under
consideration we have to find another representation of the algebra
given in Table~\ref{table}.  This representation is constructed so
that all the operators which are not constraints (\ref{nonconstr})
contain an arbitrary parameter or be zero in new representation.
To built the new representation we enlarge the representation space by
introducing the new creation and annihilation operators:  two pairs of
bosonic $b_1^+$, $b_1$ and $b_2^+$, $b_2$ and one pair of fermionic
$d^+$, $d$ creation and annihilation operators which satisfy the
standard commutation relations
\begin{eqnarray}
\bigl\{d,d^+\bigr\}=1,
\qquad
[\,b_1,b_1^+]= [\,b_2,b_2^+]= 1.
\end{eqnarray}
Then
construction of the new representation consists in extending of the
operator expressions (\ref{T0}), (\ref{T1}), (\ref{L}), (\ref{L+}),
(\ref{nonconstr}) with the help of the additional creation and
annihilation operators so that the new expressions of the operators
have the desired properties.  We will use the following new
representation for the algebra
\begin{eqnarray}
T_{0new}&=&T_0=\tilde{\gamma}^\mu p_\mu+\tilde{\gamma}m, \\
T_{1new}&=&\tilde{\gamma}^\mu a_\mu-\tilde{\gamma}b_1
-d^+b_2-2(b_2^+b_2+h)d, \label{T1new} \\
T_{1new}^+&=&\tilde{\gamma}^\mu a_\mu^+
     -\tilde{\gamma}b_1^+ + 2b_2^+d+d^+,
\label{T1+new}
\\
L_{0new}&=&L_0=-p^2+m^2,
\\
L_{1new}&=&a^{\mu}p_\mu+mb_1
\\
L_{1new}^+&=&a^{+\mu} p_\mu + mb_1^+,
\\
L_{2new}&=&{\textstyle\frac{1}{2}}\,a_\mu a^\mu-\frac{1}{2}b_1^2
   +(d^+d+b_2^+b_2+h)b_2.
\label{L2new}
\\
L_{2new}^+&=&{\textstyle\frac{1}{2}}\,a_\mu^+a^{\mu+}
   -{\textstyle\frac{1}{2}}b_1^{+2}+b_2^+,
\label{L2+new}
\\
G_{0new}&=&-a_\mu^+a^\mu
   +{\textstyle\frac{D}{2}}+d^+d+b_1^+b_1+2b_2^+b_2
   +h+{\textstyle\frac{1}{2}},
\\
M_{1new}&=&0,
\\
M_{2new}&=&0.
\end{eqnarray}
Thus we see that the operator $G_0$ which is not a constraint
contain an arbitrary parameter $h$, and the two other
operators-nonconstraints $M_1$ and $M_2$ are zero.
It may be shown that this new representation of the algebra can
be obtained from the new representation of the corresponding
massless operator algebra \cite{0410215} by dimensional
reduction $R^{1,D}\to{}R^{1,D-1}$
with the following decomposition
\begin{eqnarray}
&&
p_M=(p_\mu,m),
\qquad
a^M=(a^\mu,b_1),
\qquad
a^{+M}=(a^{+\mu},b_1^+),
\qquad
\gamma^M=(\tilde{\gamma}^\mu,\tilde{\gamma}),
\\
&&
M=0,1,\ldots,D,
\qquad
\mu=0,1,\ldots,D-1,
\qquad
\eta^{DD}=-1
.
\end{eqnarray}

It is easy to see, the operators
(\ref{T1new}), (\ref{T1+new}) and (\ref{L2new}), (\ref{L2+new})
are not Hermitian conjugate to each other
\begin{eqnarray}
T_{1new}^+\neq (T_{1new})^+,
&\qquad&
L_{2new}^+\neq (L_{2new})^+
\end{eqnarray}
if we use the usual rules for Hermitian conjugation of the
additional creation and annihilation operators
\begin{eqnarray}
d^+=(d)^+,&\qquad&b_2^+=(b_2)^+.
\end{eqnarray}
To make them conjugate each other we change the scalar product
in the enlarged space
$\langle\tilde{\Psi}_1|\Psi_2\rangle_{new}=\langle\tilde{\Psi}_1|K_h|\Psi_2\rangle$
with
\begin{eqnarray}
\label{K}
K_h&=&
\sum_{n=0}^\infty\frac{1}{n!}
  \Bigl(\,
     |n\rangle{}\langle{}n|\,C(n,h)
     -
     2d^+|n\rangle\langle{}n|d\,C(n+1,h)\,
  \,\Bigr),
\\&&
C(n,h)=h(h+1)\cdots(h+n-1),
\qquad
C(0,h)=1,
\qquad
|n\rangle=(b_2^+)^n|0\rangle
.
\end{eqnarray}
As a result we get that the operators
(\ref{T1new}), (\ref{T1+new}) and (\ref{L2new}), (\ref{L2+new})
be conjugate each other relatively the new scalar product since
the following relation take place
\begin{eqnarray}
K_h T_{1new}=(T_{1new}^+)^+K_h,
&\qquad&
K_h T_{1new}^+=(T_{1new})^+K_h,
\\
K_h L_{2new}=(L_{2new}^+)^+K_h,
&\qquad&
K_h L_{2new}^+=(L_{2new})^+K_h.
\end{eqnarray}

Thus we have constructed the representation of the algebra given
in Table~\ref{table} which possesses the properties formulated in the
beginning of this section, i.e., the operators-nonconstraints
(\ref{nonconstr}) are zeros or contain an arbitrary parameter. Then
the BRST operator will reproduce the proper equations (\ref{01'}) or
(\ref{01}) up to gauge transformation.

\section{Lagrangian for the free massive fermionic fields}\label{Lagr}

Now we construct Lagrangian for free massive
higher spin fermionic fields.
For this we construct BRST operator as if all the operators of
the algebra are the first class constraints
\begin{eqnarray}
\nonumber
\tilde{Q}
&=&
q_0T_0+q_1^+T_{1new}+q_1T_{1new}^+
+\eta_0L_0+\eta_1^+L_{1new}+\eta_1L_{1new}^+
+\eta_2^+L_{2new}
+\eta_2L_{2new}^+
\\&&
\nonumber
\qquad{}
+\eta_{G}G_{0new}
+i(\eta_1^+q_1-\eta_1q_1^+)p_0
-i(\eta_Gq_1+\eta_2q_1^+)p_1^+
+i(\eta_Gq_1^++\eta_2^+q_1)p_1
\\&&
\nonumber{}
+(q_0^2-\eta_1^+\eta_1){\cal{}P}_0
+(2q_1q_1^+-\eta_2^+\eta_2){\cal{}P}_G
+(\eta_G\eta_1^++\eta_2^+\eta_1-2q_0q_1^+){\cal{}P}_1
\\&&{}
+(\eta_1\eta_G+\eta_1^+\eta_2-2q_0q_1){\cal{}P}_1^+
+2(\eta_G\eta_2^+-q_1^{+2}){\cal{}P}_2
+2(\eta_2\eta_G-q_1^2){\cal{}P}_2^+
\label{auxBRST}
\end{eqnarray}
and assume that the
state vectors $|\chi\rangle$ and the gauge parameters
in the enlarged space (including ghosts) are
independent of the ghosts corresponding the operators which are
not constraints
\begin{eqnarray}
|\chi\rangle
&=&
\sum_{k_i}(q_0)^{k_1} (q_1^+)^{k_2} (p_1^+)^{k_3} (\eta_0)^{k_4}
(d^+)^{k_5}(\eta_1^+)^{k_6} ({\cal{}P}_1^+)^{k_7}
(\eta_2^+)^{k_8}
({\cal{}P}_2^+)^{k_9}(b_1^+)^{k_{10}}(b_2^+)^{k_{11}}
\times
\nonumber
\\
&&\qquad{}
\times
a^{+\mu_1}\cdots a^{+\mu_{k_0}}
\chi_{\mu_1\cdots\mu_{k_0}}^{k_1\cdots{}k_{11}}(x)|0\rangle.
\label{chistate}
\end{eqnarray}
The corresponding ghost number of this vector is $0$, as usual.
The sum in (\ref{chistate}) is assumed over $k_0$, $k_1$, $k_2$,
$k_3$, $k_{10}$, $k_{11}$ running from 0 to infinity and over $k_4$, $k_5$, $k_6$,
$k_7$, $k_8$, $k_9$ running from 0 to 1.
Let us notice that the new BRST charge (\ref{auxBRST}) is
selfconjugate in the following sense
$\tilde{Q}^+K_h=K_h\tilde{Q}$,
with operator $K_h$ (\ref{K}).

Lagrangian for the massive fermionic field with spin
$s=n+1/2$ is constructed as follows
(see details in \cite{0410215})\footnote{The Lagrangian is defined as usual
up to an overall factor}
\begin{eqnarray} {\cal{}L}_n
&=&
{}_n\langle\tilde{\chi}{}^{0}_{0}|K_n\tilde{T}_0|\chi^{0}_{0}\rangle_n
+
\frac{1}{2}\,{}_n\langle\tilde{\chi}{}^{1}_{0}|K_n\bigl\{
   \tilde{T}_0,\eta_1^+\eta_1\bigr\}|\chi^{1}_{0}\rangle_n
\nonumber
\\&&\qquad{}
+
{}_n\langle\tilde{\chi}{}^{0}_{0}|K_n\Delta{}Q_{n}|\chi^{1}_{0}\rangle_n
+
{}_n\langle\tilde{\chi}{}^{1}_{0}|K_n\Delta{}Q_{n}|\chi^{0}_{0}\rangle_n
,
\label{L1}
\end{eqnarray}
where $\tilde{T}_0=T_0-2q_1^+{\cal{}P}_1-2q_1{\cal{}P}_1^+$ and
$\{A,B\}=AB+BA$.
In eq.~(\ref{L1}) $|\chi^0_0\rangle_n$, $|\chi^1_0\rangle_n$ are coefficients of
$|\chi\rangle$ (\ref{chistate}) standing at
$(q_0)^0(\eta_0)^0$ and $(q_0)^1(\eta_0)^0$ respectively which
subject to the condition
\begin{eqnarray}
\sigma|\chi^i_0\rangle_n
&=&
\bigl(n+{\textstyle\frac{D-3}{2}}\bigr)|\chi^i_0\rangle_n,
\qquad
i=0,1,
\qquad
gh(|\chi^i_0\rangle_n)=-i,
\label{chicond}
\\
\sigma
&=&
-a_\mu^+a^\mu+{\textstyle\frac{D+1}{2}}+d^+d+b_1^+b_1+2b_2^+b_2
-iq_1p_1^++iq_1^+p_1
\nonumber
\\
&&\qquad{}
+\eta_1^+{\cal{}P}_1-\eta_1{\cal{}P}_1^+
+2\eta_2^+{\cal{}P}_2-2\eta_2{\cal{}P}_2^+
,
\end{eqnarray}
$\Delta{}Q_n$ is the part of $\tilde{Q}$ (\ref{auxBRST}) which
independent of the ghosts $\eta_G$, ${\cal{}P}_G$, $\eta_0$,
${\cal{}P}_0$, $q_0$, $p_0$ and the substitution
$-h\to{}n+(D-3)/2$ is done, $K_n$ is operator (\ref{K}) where
the substitution $-h\to{}n+(D-3)/2$ is done.

Lagrangian (\ref{L1}) is invariant under the gauge
transformation
\begin{eqnarray}
\delta|\chi^{0}_{0}\rangle_n
&=&
\Delta{}Q_{n}|\Lambda^{0}_{0}\rangle_n
 +
 \frac{1}{2}\bigl\{\tilde{T}_0,\eta_1^+\eta_1\bigr\}
 |\Lambda^{1}_{0}\rangle_n,
\label{GT1}
\\
\delta|\chi^{1}_{0}\rangle_n
&=&
\tilde{T}_0|\Lambda^{0}_{0}\rangle_n
 +\Delta{}Q_{n}|\Lambda^{1}_{0}\rangle_n
 .
\label{GT2}
\end{eqnarray}
which are reducible
\begin{align}
\delta|\Lambda^{(i)}{}^{0}_{0}\rangle_n
&=
\Delta{}Q_{n}|\Lambda^{(i+1)}{}^{0}_{0}\rangle_n
 +
 \frac{1}{2}\bigl\{\tilde{T}_0,\eta_1^+\eta_1\bigr\}
 |\Lambda^{(i+1)}{}^{1}_{0}\rangle_n,
&
|\Lambda^{(0)}{}^0_0\rangle_n=|\Lambda^0_0\rangle_n,
\label{GTi1}
\\
\delta|\Lambda^{(i)}{}^{1}_{0}\rangle_n
&=
\tilde{T}_0|\Lambda^{(i+1)}{}^{0}_{0}\rangle_n
 +\Delta{}Q_{n}|\Lambda^{(i+1)}{}^{1}_{0}\rangle_n,
&
|\Lambda^{(0)}{}^1_0\rangle_n=|\Lambda^1_0\rangle_n,
\label{GTi2}
\end{align}
with finite number of reducibility stages $i_{max}=n-1$ for spin
$s=n+1/2$. In the above formulae the gauge parameters
$|\Lambda^{(k)}{}^i_0\rangle_n$ are subject to the conditions
which are analogous to the conditions on $|\chi^0_0\rangle_n$,
$|\chi^1_0\rangle_n$. Namely, the gauge parameters are independent
on the ghosts $\eta_G$, ${\cal{}P}_G$, $\eta_0$, ${\cal{}P}_0$,
$q_0$, $p_0$ and the following conditions are fulfilled
\begin{eqnarray}
\sigma|\Lambda^{(k)}{}^i_0\rangle_n
&=&
\bigl(n+{\textstyle\frac{D-3}{2}}\bigr)
|\Lambda^{(k)}{}^i_0\rangle_n,
\qquad
gh(|\Lambda^{(k)}{}^i_0\rangle_n)=-(i+k+1).
\label{lambdacond}
\end{eqnarray}

Let us show that Lagrangian (\ref{L1}) reproduces the correct
equations (\ref{01}) on the physical field. As we already mention
the massive theory looks like dimensional reduction of the
corresponding massless theory which presented in \cite{0410215}.
First of all we remove all the
auxiliary fields associated with the ghost fields and associated
with the additional creation operators $d^+$, $b_2^+$ analogously
to Section~6 of \cite{0410215}. After that the only auxiliary
fields we get are ones associated with operator $b_1^+$. After
this we get the equations of motion
\begin{eqnarray}
(T_0+\tilde{L}_1^{+}\tilde{T}_1)|\Psi\rangle_n=0, \qquad
(\tilde{T}_1)^3|\Psi\rangle_n=0,
\label{em}
\end{eqnarray}
which are invariant under the gauge transformation
\begin{eqnarray}
\delta|\Psi\rangle_n=\tilde{L}_1^+|\Lambda\rangle_{n-1},
\qquad
\tilde{T}_1|\Lambda\rangle_{n-1}=0.
\label{gt}
\end{eqnarray}
Here $T_0$ is given by (\ref{T0}) and
\begin{eqnarray}
&&
|\Psi\rangle_n
=
\sum_{k=0}^{n}(b_1^+)^k a^{+\mu_1}\ldots a^{+\mu_{n-k}}
\psi_{\mu_1\ldots\mu_{n-k}}(x)|0\rangle,
\label{PSI}
\\
&&
|\Lambda\rangle_{n-1}
=
\sum_{k=0}^{n-1}(b_1^+)^k a^{+\mu_1}\ldots a^{+\mu_{n-k-1}}
\lambda_{\mu_1\ldots\mu_{n-k-1}}(x)|0\rangle,
\label{L-bda}
\\
&&
\tilde{L}_1^+=
a^{+\mu}p_\mu+mb_1^+,
\qquad
\tilde{T}_1=\tilde{\gamma}^\mu a_\mu-\tilde{\gamma}b_1.
\end{eqnarray}
Now we get rid of the auxiliary fields associated with $b_1^+$.
First we note that
the second equation in (\ref{gt}) tells us that
$\lambda_{\mu_1\ldots\mu_{n-1}}$ in (\ref{L-bda}) is unconstrained
and the other coefficients in (\ref{L-bda})
$\lambda_{\mu_1\ldots\mu_{n-k-1}}$ are expressed through it
$\lambda_{\mu_1\ldots\mu_{n-k-1}}
\sim
\tilde{\gamma}^{\mu_1}\ldots\tilde{\gamma}^{\mu_k}
\lambda_{\mu_1\ldots\mu_{n-1}}$.
Then we decompose equation~(\ref{em}) in power series of
$b_1^+$ and
using the on-shell gauge transformations we remove
first $\psi$ with the help of
$\tilde{\gamma}^{\mu_1}\ldots\tilde{\gamma}^{\mu_{n-1}}
\lambda_{\mu_1\ldots\mu_{n-1}}$,
then $\psi_\mu$ with the help of
$\tilde{\gamma}^{\mu_1}\ldots\tilde{\gamma}^{\mu_{n-2}}
\lambda_{\mu_1\ldots\mu_{n-1}}$,
and so on till
$\psi_{\mu_1\ldots\mu_{n-1}}$.
Thus we removed all the auxiliary fields
and
after this we find that the physical field
$\psi_{\mu_1\ldots\mu_{n}}$ obey the equations
(\ref{irrep0}), (\ref{irrep1}).

Now we turn to examples.

\section{Examples}\label{example}

\subsection{Spin 1/2}
We begin with the simplest case of spin $s=1/2$ and show that Lagrangian
(\ref{L1}) reproduces Dirac Lagrangian.  In
this case there exists the field $|\chi^0_0\rangle$ only which
obeys condition (\ref{chicond})
\begin{eqnarray}
|\chi^0_0\rangle_0=\psi(x)|0\rangle, \qquad
{}_0\langle\tilde{\chi}{}^0_0|=\langle0|\psi^+\tilde{\gamma}{}^0.
\label{x00}
\end{eqnarray}
Substituting (\ref{x00}) in (\ref{L1}) ones get
\begin{eqnarray}
{\cal{}L}_{0}
&=&
{}_0\langle\tilde{\chi}{}^0_0|K_0T_0|\tilde{\chi}{}^0_0\rangle_0
=-\bar{\psi}\bigl(i\gamma^\mu\partial_\mu-m\bigr)\psi.
\end{eqnarray}
Here we used the definition for the "true" gamma-matrices
(\ref{truegamma}) and introduced the Dirac conjugate spinor
\begin{eqnarray}
\bar{\psi}&=&\psi^+\gamma^0.
\label{Dirac}
\end{eqnarray}
Thus we see that Lagrangian (\ref{L1}) reproduces the Dirac
Lagrangian up to an overall factor.

\subsection{Spin 3/2}

Our aim is to construct Lagrangian in
terms of "true" gamma-matrices (\ref{truegamma}).
To do this we assign a definite Grassmann parity to the
fields and the gauge parameters.
For example we can take the field $|\chi\rangle$ to be
even (odd) then the gauge parameter
$|\Lambda^{(0)}\rangle$ will be odd (even), then the
gauge parameter $|\Lambda^{(1)}\rangle$ will be even
(odd), and so on.
For this purpose we will use $\tilde{\gamma}$
to keep the proper Grassmann parity of the fields
$|\chi^i_0\rangle$ and the gauge parameters
$|\Lambda^{(k)}{}^i_0\rangle$.

In the case of
$s=3/2$ we have the following expressions for the fields
$|\chi^0_0\rangle$, $|\chi^1_0\rangle$ and the gauge parameter
$|\Lambda^0_0\rangle$ which obey the above Grassmann parity conditions and conditions
(\ref{chicond}) and (\ref{lambdacond}) respectively
\begin{eqnarray} |\chi^0_0\rangle_1 &=&
\bigl[-ia^{+\mu}\psi_\mu(x)+d^+\tilde{\gamma}\psi(x)
  +b_1^+\varphi(x)\bigr]
|0\rangle
,
\label{chi001}
\\
\langle\tilde{\chi}{}^0_0|
&=&\langle0|\bigl[
\psi_\mu^+(x)ia^{+\mu}+\psi^+(x)\tilde{\gamma}d
+\varphi^+(x)b_1
\bigr]\tilde{\gamma}{}^0
,
\\
|\chi^1_0\rangle_1
&=&
\bigl[{\cal{}P}_1^+\tilde{\gamma}\chi(x)
 -ip_1^+\chi_1(x)\bigr]|0\rangle
,
\label{chi101}
\\
\langle\tilde{\chi}{}^1_0|
&=&
\langle0|\bigl[\chi_1^+(x)ip_1
+\chi^+(x)\tilde{\gamma}{\cal{}P}_1\Bigr]\tilde{\gamma}{}^0
,
\label{<chi101}
\\
|\Lambda^0_0\rangle_1
&=&
\bigl[{\cal{}P}_1^+\lambda(x)
 -ip_1^+\tilde{\gamma}\lambda_1(x)\bigr]|0\rangle
.
\label{l001}
\end{eqnarray}
Substituting (\ref{chi001})--(\ref{<chi101}) into (\ref{L1}) ones
find Lagrangian (up to an overall factor) for spin-3/2 field
\begin{eqnarray}
{\cal{}L}_1&=&
\bar{\psi}{}^\mu\Bigl[
(i\gamma^\sigma\partial_\sigma-m)\psi_\mu-i\gamma_\mu\chi_1
-\partial_\mu\chi
\Bigr]
+
\bar{\varphi}\Bigl[
(-i\gamma^\mu\partial_\mu+m)\varphi+m\chi-\chi_1
\Bigr]
\nonumber
\\
&&{}
+(D-1)\bar{\psi}\Bigl[
(i\gamma^\mu\partial_\mu+m)\psi+\chi_1
\Bigr]
+
\bar{\chi}\Bigl[
(i\gamma^\mu\partial_\mu+m)\chi-\chi_1
+\partial^\mu\psi_\mu+m\varphi
\Bigr]
\nonumber
\\
&&{}
+\bar{\chi}_1\Bigl[
-\chi+i\gamma^\mu\psi_\mu-\varphi+(D-1)\psi
\Bigr]
\label{L3/2}
\end{eqnarray}
and substituting (\ref{chi001}), (\ref{chi101}), (\ref{l001})
into (\ref{GT1}),
(\ref{GT2}) ones find gauge transformations for the fields
\begin{eqnarray}
\delta \psi_\mu
&=&
\partial_\mu\lambda+i\gamma_\mu\lambda_1
,
\qquad
\delta \psi=\lambda_1,
\qquad
\delta\varphi=m\lambda+\lambda_1,
\\
\delta\chi
&=&
(i\gamma^\mu\partial_\mu-m)\lambda
-2\lambda_1,
\\
\delta \chi_1
&=&
-(i\gamma^\mu\partial_\mu+m)\lambda_1.
\end{eqnarray}
Here we have used (\ref{truegamma}) and (\ref{Dirac}) and that
$K_hd^+|0\rangle=-2hd^+|0\rangle$ with $-2h\to(D-1)$.

One can easy check that removing $\varphi$ and $\psi$ with the
help of their gauge transformation using the gauge parameters
$\lambda$ and $\lambda_1$ respectively we get that $\chi_1=0$ and
then $\chi=0$ as consequences of the equations of motion.  After
this the equations on the physical field $\psi_\mu$ coincide with
the equations which define the irreducible representation with
spin $s=3/2$.

Now we transform (\ref{L3/2}) to Rarita-Schwinger Lagrangian. For this
purpose we remove $\varphi$ and $\psi$ with the help of their gauge
transformations using the parameters $\lambda$ and $\lambda_1$
respectively.  After this we integrate over $\chi_1$ and then over
$\chi$ (or the same we use algebraic equation of motion
$\chi=i\gamma^\mu\psi_\mu$~) and get
\begin{eqnarray} {\cal{}L}_{RS}&=&
\bar{\psi}{}^\mu(i\gamma^\sigma\partial_\sigma-m)\psi_\mu
-i\bar{\psi}{}^\mu
 (\gamma^\nu\partial_\mu+\gamma_\mu\partial^\nu)
 \psi_\nu
+
\bar{\psi}{}_\mu\gamma^\mu
 (i\gamma^\sigma\partial_\sigma+m)\gamma^\nu\psi_\nu
.
\end{eqnarray}
This is the Rarita-Schwinger Lagrangian in $D$
dimensions.

\section{Summary}\label{Summary}

We have developed the operator approach to derivation of gauge
invariant Lagrangians for fermionic massive higher spin models in
arbitrary dimensional Minkowski space. The approach is based on
realization of the conditions determining the fermionic higher
spin fields as the operator constraints acting in auxiliary Fock
space, finding the closed algebra of the constraints and
construction of the BRST operator (or to be more precise, BRST-BFV
operator \cite{BRST}) in this space.

We found that the model under consideration is a reducible gauge
theory and the order of reducibility linearly grows with the value
of spin.  It is shown that the BRST operator generates the
consistent Lagrangian dynamics for fermionic fields of any value
of spin in space of arbitrary dimension. As an example of general
scheme we obtained the gauge invariant Lagrangian and the gauge
transformations for the massive fields with spin-3/2 in the
explicit form. Derivation of gauge invariant Lagrangians for any
other spins on the base of our general approach is a purely
technical problem.

The main results of the paper are given by the relations
(\ref{L1}), where Lagrangian for the field with arbitrary
half-integer spin is given, and (\ref{GT1})--(\ref{GTi2}) where
the gauge transformations for the fields and the gauge parameters
are written down. The formulation does not assume to impose any
off-shell constraints on the fields and the gauge parameters from
the very beginning.  All the constraints emerge as the
consequences of the equations of motion and gauge fixing. The
approach can also be applied to Lagrangian construction for
fermionic higher spin fields in AdS space and for the fermionic
fields with mixed symmetry.

\section*{Acknowledgements}
Work of I.L.B and V.A.K was supported in part by the
INTAS grant, project INTAS-03-51-6346,
the RFBR grant, project No.\ 06-02-16346 and grant for LRSS, project
No.\ 4489.2006.2. Work of L.L.R was partially supported by the grant
for LRSS, project No.\ 4489.2006.2.  Work of I.L.B was supported in part
by the DFG grant, project No.\ 436 RUS 113/669/0-3.

\end{document}